\documentclass[aps,pra,reprint,superscriptaddress]{revtex4-2}

\usepackage{graphicx}
\usepackage{amsmath}
\usepackage{braket}
\usepackage{xcolor}
\usepackage[normalem]{ulem}
\usepackage{notes2bib}
\bibnotesetup{
note-name = ,
use-sort-key = false
}
\newcommand{\expect}[1]{\langle#1\rangle}

\usepackage{hyperref}
\hypersetup{
    colorlinks=true,
    citecolor=blue,
    linkcolor=blue,
    filecolor=blue,      
    urlcolor=blue,
}
\usepackage[nameinlink,capitalise]{cleveref}

\begin{document}

\title{Autoregressive neural quantum states of Fermi Hubbard models}

\author{Eduardo Ibarra-Garc\'ia-Padilla}
\email[]{edibarra@ucdavis.edu}
\affiliation{Department of Physics and Astronomy, University of California, Davis, CA 95616, USA}
\affiliation{Department of Physics and Astronomy, San Jos\'e State University, San Jos\'e, CA 95192, USA}
\author{Hannah Lange}
\affiliation{Ludwig-Maximilians-University Munich, Theresienstr. 37, Munich D-80333, Germany}
\affiliation{Max-Planck-Institute for Quantum Optics, Hans-Kopfermann-Str.1, Garching D-85748, Germany}
\affiliation{Munich Center for Quantum Science and Technology, Schellingstr. 4, Munich D-80799, Germany}
\author{Roger G Melko}
\affiliation{Department of Physics and Astronomy, University of Waterloo, 200 University Ave. West, Waterloo, Ontario N2L 3G1, Canada}
\affiliation{Perimeter Institute for Theoretical Physics, Waterloo, Ontario N2L 2Y5, Canada}
\author{Richard T Scalettar}
\affiliation{Department of Physics and Astronomy, University of California, Davis, CA 95616, USA}
\author{Juan Carrasquilla}
\affiliation{Institut f\"ur Theoretische Physik, Eidgen\"ossische Technische Hochschule Z\"urich, Wolfgang-Pauli-Strasse 27, 8093 Z\"urich, Switzerland}
\author{Annabelle Bohrdt}
\affiliation{Munich Center for Quantum Science and Technology, Schellingstr. 4, Munich D-80799, Germany}
\affiliation{University of Regensburg, Universit\"atsstr. 31, Regensburg D-93053, Germany}
\author{Ehsan Khatami}
\email[]{ehsan.khatami@sjsu.edu }
\affiliation{Department of Physics and Astronomy, San Jos\'e State University, San Jos\'e, CA 95192, USA}

\date{\today}

\begin{abstract}
Neural quantum states (NQS) have emerged as a powerful ansatz for variational quantum Monte Carlo studies of strongly-correlated systems. Here, we apply recurrent neural networks (RNNs) and autoregressive transformer neural networks to the Fermi-Hubbard and the (non-Hermitian) Hatano-Nelson-Hubbard models in one and two dimensions. In both cases, we observe that the convergence of the RNN ansatz is challenged when increasing the interaction strength. We present a physically-motivated and easy-to-implement strategy for improving the optimization, namely, by ramping of the model parameters. Furthermore, we investigate the advantages and disadvantages of the autoregressive sampling property of both network architectures. For the Hatano-Nelson-Hubbard model, we identify convergence issues that stem from the autoregressive sampling scheme in combination with the non-Hermitian nature of the model. Our findings provide insights into the challenges of the NQS approach and make the first step towards exploring strongly-correlated electrons using this ansatz.

\end{abstract}

\maketitle

\section{Introduction}
 
The Fermi-Hubbard model (FHM) describes itinerant, interacting spin-$1/2$ electrons hopping on a set of spatially localized orbitals. Despite its simplicity, it is paradigmatic for our understanding of electronic correlations in quantum materials in which strong Coulomb interactions play an essential role. Its importance lies in the fact that it accurately captures some of the key characteristics of strongly correlated materials. Numerous phases this model can display have striking similarities to the behaviors observed in a wide range of complex materials~\cite{rasetti1991hubbard,gebhard1997metal,fazekas1999lecture,Arovas2021,Qin2021}.

In recent years, neural network quantum states (NQS)~\cite{Carleo2017,g_torlai_17,r_melko_19,z_jia_19,j_carrasquilla_20,j_carrasquilla_21,Bukov2021,Medvidovic2024,Lange_2024,Reh2023,Yoshioka2019} have emerged as promising ansatz for variational wave functions. There are several remarkable aspects that make NQS attractive for use in the field of quantum many-body physics. First is the ability of the NQS to capture a wide array of quantum states important to the study of quantum many-body systems~\cite{Medvidovic2024,Lange_2024}. It has been shown that they can encode volume law entangled states \cite{Sharir2022,Deng2017,Gao2017,denis2023comment,Levine2019}. 

The second feature is related to the scaling of computational resources with the system size. The optimization and evaluation of observables in NQS relies on sampling, typically using a Metropolis sampling algorithm, which can become computationally very demanding. Here, we show that {\it autoregressive} networks, with normalized wave function amplitudes that allow direct sampling instead of Metropolis sampling, can be very efficient for correlated electron systems. The autoregressive property refers to the use of a chain rule of probabilities to generate a sequence of data elements in which the probability of every element in the sequence depends only on the configuration of the set of elements that came before it. The scaling of the method with the system size can lead to a potentially huge advantage over conventional numerical treatment, such as the exact diagonalization (ED), density matrix renormalization group (DMRG), or quantum Monte Carlo (QMC)~\cite{DMRG_White,DMRG_review,DQMC_1,DQMC_2}, which suffer from either exponential scaling or other technical issues arising due to the fermion `sign problem'~\cite{Loh1989,Troyer2005,v_iglovikov_15}.

Two prominent examples for autoregressive networks are recurrent neural networks (RNNs)~\cite{Hibat2020} and transformer neural networks~\cite{Luo2022,Luo2023,zhang_transformer_2023,sprague2023variational,Lange2024,fitzek2024} (although non-autoregressive versions of transformers are possible, see e.g. \cite{viteritti_transformer_2023,rende_simple_2023}). 
RNNs were initially developed for natural language processing and are {\it generative}, i.e., they can be trained to infer the probability distribution of unlabeled data, which can then be used to generate more data. 
The final joint distribution of configurations in an RNN, $P(\sigma)$, where $\sigma$ represents a configuration in a complete set, is normalized. This is a powerful property; not only can one sample from a trained RNN, but also given a new configuration, the trained RNN can return the associated normalized probability. The same property can be enforced for a transformer neural network by masking out future inputs to the attention mechanism, see e.g. Refs. \cite{Luo2022,Luo2023,zhang_transformer_2023,sprague2023variational,Lange2024,fitzek2024}.

In a pioneering work~\cite{Hibat2020}, Hibat-Allah {\it et.~al.}~showed that ground states of spin models can accurately be represented using a wave function based on an RNN with two output layers for amplitude and phase of the wave function. 
Namely, samples are drawn from the generative model, the energy is computed and averaged over many samples, and minimized with respect to the parameters of the RNN using the stochastic gradient descent. Weights and biases inside a shared RNN unit are among the quantities that form the optimization parameters. 
Since then, there have been several notable studies exploring aspects of the use of RNNs in quantum many-body physics, spanning representability, accuracy, performance~\cite{s_czischek_22,m_hibatallah_23,Moss2024,h_lange_23,Doeschl2023}, including through the use of symmetries of the Hamiltonian~\cite{s_morawetz_20} or more advanced autoregressive architectures such as {\it transformers}~\cite{bennewitzNeuralErrorMitigation2022,zhang_transformer_2023,sprague2023variational,Lange2024}. 

Neural network wavefunctions have also been used recently to study itinerant electron models at half-filling and beyond ~\cite{Stokes2020,Nomura2017,liu2023unifying,RobledoMoreno2022,Luo2019,Hermann2020,pfau2020ferminet,kim2023neuralnetwork,romero2024spectroscopy,Humeniuk2022,Inui2021,Yoshioka2021,h_lange_23,Bortone2024}. In Ref.~\cite{h_lange_23}, the authors use RNNs to represent the ground state of the $t-J$, $t-{XXZ}$, and $t-{J_z}$ models in one and two spatial dimensions. They present a novel technique for mapping out the dispersion relation of the single hole, complementary to other schemes \cite{Choo2018,Nomura2020disp,Viteritti2022accuarcy}, and show that their results compare well with DMRG. This is accomplished through computing the expectation values of the translation operator for periodic systems. However, they establish that the ground state uncertainties for these models are generally much larger than those for spin models studied previously.

Here, we utilize RNNs as a variational ansatz to access the ground state of the FHM. We find that, as expected, a naive application of the technique results in accuracies that are less accurate than in the case of the $t-J$ model. We then introduce a ramping mechanism, in which certain model parameters are gradually tuned from the initial values to the desired final values during the training process, and show that the scheme leads to orders of magnitude improvements in the ground state energy and other observables (this is also known as variational neural annealing~\cite{variational_neural_annealing}).
We benchmark our results for the method applied to the one-dimensional (1D) and two-dimensional (2D) FHM at half filling and find that the accuracy of ground state properties is generally independent of the system size for the same number of training steps, confirming that the computational resources will indeed grow linearly with system size. We then use the method to study a non-Hermitian Hamiltonian in which the tunneling rates to the left and right are unequal [the Hatano-Nelson-Hubbard model (HNHM)], and discuss the limitations of current RNN architectures for such problems.

The remainder of this paper is organized as follows. In Sec.~\ref{sec::Model_Methods} we present the Fermi-Hubbard model, the observables computed, the details of the RNNs, and the training schemes used. In Sec.~\ref{sec::Results} we present our main findings, first for the 1D FHM, second for the 2D FHM, and third for the 1D HNHM. Section~\ref{sec::Conc} summarizes our findings and presents an outlook for future studies.

\section{Model and methods}\label{sec::Model_Methods} 

\subsection{Fermi-Hubbard model}
We study the FHM, whose Hamiltonian is expressed as
\begin{equation}\label{eq:Hubbard_N1}
\hat{H}_{\mathrm{FH}} = -t \sum_{\langle i,j \rangle, \sigma} \left( \hat{c}_{i \sigma}^\dagger \hat{c}_{j \sigma}^{\phantom{\dagger}} 
+ \mathrm{h.c.} \right) + U \sum_{i} \hat{n}_{i \uparrow} \hat{n}_{i \downarrow}  - \mu \sum_{i,\sigma} \hat{n}_{i \sigma},
\end{equation} 
where $\hat{c}_{i \sigma}^\dagger$ ($\hat{c}_{i \sigma}^{\phantom{\dagger}} $) is the creation (annihilation) operator for a fermion with spin $\sigma$ on site $i$, $\hat{n}_{i \sigma} = \hat{c}_{i \sigma}^\dagger \hat{c}_{i \sigma}^{\phantom{\dagger}}$ is the number operator for spin $\sigma$ on site $i$, $\expect{i,j}$ denotes sum over nearest neighbors, $t$ is the nearest-neighbor hopping amplitude, $U$ is the interaction strength, and $\mu$ is the chemical potential that controls the fermion density in the grand canonical ensemble. We consider the repulsive case $U>0$. We set the energy scale to be $t=1$. We consider 1D chains with $N=L$ sites and 2D square lattices with $N = L_x \times L_y$ sites. In all cases, open boundary conditions (OBC) are considered.
We work in the grand canonical ensemble and set the chemical potential to $\mu =U/2$ to achieve half-filling on average in all cases involving this model.

The HNHM is described by a non-Hermitian Hamiltonian with unequal tunneling rates for left and right directions,
\begin{align}\label{eq:HNH}
\hat{H}_{\mathrm{HNH}} &= -t \sum_{i, \sigma} \bigg[ \left(1+g/t \right) \hat{c}_{i+1 \sigma}^\dagger \hat{c}_{i \sigma}^{\phantom{\dagger}} 
+ \left(1-g/t\right)\hat{c}_{i\sigma}^\dagger \hat{c}_{i+1\sigma}^{\phantom{\dagger}}  \bigg] \nonumber \\
&+ U \sum_{i} \hat{n}_{i \uparrow} \hat{n}_{i \downarrow}  - \mu \sum_{i,\sigma} \hat{n}_{i \sigma}.
\end{align}

An important consequence of the anisotropic tunneling rates is that the HNHM is particle-hole symmetric (PHS) under the transformation $\hat{c}_{i\sigma}^\dagger \to (-1)^i\hat{c}_{i\sigma}^{\phantom{\dagger}}$ and $g \to -g$. The regular particle hole transformation [only $\hat{c}_{i\sigma}^\dagger \to (-1)^i\hat{c}_{i\sigma}^{\phantom{\dagger}}$] maps the left kinetic energy term into the right one, and vice-versa, therefore the change in sign of $g$ is crucial to achieve PHS.

We compute the energy $E = \expect{\hat{H}}/N$, the density
\begin{equation}
    n = \frac{1}{N} \sum_{i,\sigma} \expect{\hat{n}_{i\sigma}}, 
\end{equation}
the kinetic energy
\begin{equation}
    K = \frac{1}{N}\bigg\langle -t \sum_{\langle i,j \rangle, \sigma} \left(\hat{c}_{i \sigma}^\dagger \hat{c}_{j \sigma}^{\phantom{\dagger}} 
+ \mathrm{h.c.} \right) \bigg\rangle,
\end{equation}
the double occupancy
\begin{equation}
    \mathcal{D} = \frac{1}{N} \sum_i \expect{\hat{n}_{i \uparrow} \hat{n}_{i \downarrow}},
\end{equation}
and the nearest-neighbor (nn) spin-spin correlation function
\begin{align}
    \expect{S_z S_z}_{\mathrm{nn}} &= \frac{1}{N_b} \sum_{i} \sum_{\boldsymbol{\delta}\in \mathcal{S}(i)} \expect{\hat{S}_z^{i} \hat{S}_z^{i+\boldsymbol{\delta}}}, \\
    \mathrm{where} \: \expect{\hat{S}_z^{i} \hat{S}_z^{i+\boldsymbol{\delta}}} &= 
    \frac{1}{4} \sum_{\sigma} \bigg( \expect{\hat{n}_{i \sigma}\hat{n}_{i+\boldsymbol{\delta} \sigma}} - \expect{\hat{n}_{i \sigma}\hat{n}_{i+\boldsymbol{\delta} \overline{\sigma}}} \bigg) \nonumber,
\end{align}
$N_b$ is the number of bonds, $\mathcal{S}(i)$ denotes the set of nearest neighbor vectors consistent with the OBC for site $i$, and $\overline{\sigma}$ denotes the opposite spin to $\sigma$.

\subsection{Recurrent neural network wave functions}

We use a recurrent neural network (RNN) to represent a pure quantum state in each of our case studies. The wavefunction ansatz is given by
\begin{equation}\label{eq:ansatz}
    | \psi_\lambda\rangle = \sum_\sigma \sqrt{p_\lambda(\sigma)} e^{i\phi_\lambda(\sigma)} | \sigma \rangle,
\end{equation}
where $\lambda$ denotes the variational parameters of the ansatz wavefunction $| \psi_\lambda\rangle$, and $|\sigma\rangle$ are the elements of the computational basis, i.e. $|\sigma\rangle = |\sigma_1,\sigma_2,\cdots,\sigma_N\rangle$, where $\sigma_i$ can take any value of the local Hilbert space $\{0,\uparrow,\downarrow,\uparrow \downarrow\}$. $\lambda$ includes {\it hidden variables}, $h_i$, used to pass information from site $i$ to site  $i+1$ during a given sampling step. The size of $h_i$'s is given by the number of hidden units, $n_h$.

In this work, we use one RNN cell and a Softmax layer to model the probability, together with a Softsign layer to model the phase. Specifically, we implement the gated recurrent unit (GRU) as the elementary cell in our RNNs. For further details on the parametrization of the GRU and RNN in this scheme, we refer the reader to Ref.~\cite{Hibat2020}.

In order to find the ground state of the system, we perform variational Monte Carlo (VMC) to minimize the energy. In that case, we minimize the expectation value of the energy of our ansatz wavefunction
\begin{equation}
    \expect{H_\lambda} = \sum_{\sigma} |p_\lambda(\sigma)| \epsilon_\lambda(\sigma),
\end{equation}
where we have defined
\begin{equation}
    \epsilon_\lambda (\sigma) =  \sum_{\tau} H_{\tau \sigma} \sqrt{\frac{p_\lambda(\tau)}{p_\lambda(\sigma)}} e^{i[ \phi_\lambda(\tau) - \phi_\lambda(\sigma)]},
\end{equation}
and $H_{\tau \sigma} = \expect{\tau |H | \sigma}$.

In practice, we consider the cost function
\begin{equation}
    C = \sum_\sigma |p_\lambda(\sigma)|  \big[ \epsilon_\lambda(\sigma) - \expect{\epsilon_\lambda(\sigma)} \big]
\end{equation}
to minimize both the energy and its variance to stabilize the training, and we use the adaptive moment estimation (adam) optimizer to implement the gradient updates.

\subsection{Pre-training with projective measurements}

Refs.~\cite{Moss2024,s_czischek_22,Lange2024} demonstrated the potential of hybrid quantum-classical methods for large-scale quantum many-body system simulation by merging experimental data from current quantum devices with autoregressive language models. 

More specifically, they observed that if the RNN wavefunction is first trained using experimental projective measurements (which we will name pre-training) and then VMC is performed, the convergence of the procedure is significantly improved. The only difference between these two settings is the loss function governing the optimization. In the VMC stage the loss function is the energy while in the pre-training stage the loss function is the Kullback-Leibler (KL) divergence~\cite{Moss2024},
\begin{equation}
    L_\mathrm{KL} = \sum_\sigma p_d(\sigma) \ln \left[\frac{p_d (\sigma)}{p_\lambda (\sigma)}\right], 
\end{equation}
where $p_d$ is the empirical distribution of the dataset. The minimum of this loss function occurs when $p_d(\sigma) = p_\lambda(\sigma)$. Therefore, minimizing the KL divergence drives the matching of the distributions. Note that in practice, it suffices to minimize $-\langle \ln[p_\lambda (\sigma)]\rangle_{p_d}$ in this stage. 

In this work, we also experiment with pre-training our RNN models using projective measurements obtained from sampling of the ground state function in the computational basis using exact diagonalization. For each case study, we generate and use 10,000 samples.

\subsection{Ramping of Hubbard parameters}

Additionally, we propose an alternative way of enhancing VMC simulators in which Hubbard parameters are ramped to their desired value. This idea is inspired by the experimental protocol for preparing low entropy samples in optical lattices and optical tweezer arrays in which it is more efficient to load a band insulator and then modify the lattice (increase the number of sites) to get on average one particle per site or other target fillings~\cite{Chiu2018,Yan2022}. In our proposed training scheme, instead of modifying the lattice geometry (as is often done in experiments), we modify the hopping amplitude during the training.

As we will discuss in Sec.~\ref{sec::Results}, we find that ramping the tunneling rate from a larger $t$ to the desired final value provides a scheme that either alone or in conjunction with pre-training with projective measurements, provides better results than VMC with pre-training using projective measurements alone.

The reason behind starting from large $t$ rather than any of the other Hubbard parameters is three-fold: (1) While the $\mu$ and $U$ terms in the Fermi-Hubbard Hamiltonian are diagonal in the $n_{i\sigma}$ basis, the $t$ term is not. For that reason, the kinetic energy term mixes elements of the computational basis and its calculation involves knowledge of both the amplitudes and the phases (in contrast to the other two terms that only require the amplitudes). Therefore, starting from large $t$ biases the RNN to learn amplitudes and phases simultaneously. (2) Starting from a larger $t$, the system promotes double occupancies, which are essential for reaching the true ground state~\footnote{As an example, consider the two particle two site calculation, where one finds that the ground state of the model is the singlet state with a small admixture of doublons. In the $t/U \ll 1$ limit, the ground state energy is $\approx -4t^2/U$, and the ground state is given by 
\begin{equation*}
    |\psi\rangle \approx \mathcal{N} \bigg[ \frac{2t}{U} \bigg(|0,\uparrow \downarrow \rangle + |\uparrow \downarrow,0 \rangle \bigg) + \bigg(|\uparrow,\downarrow \rangle + |\downarrow,\uparrow \rangle \bigg)  \bigg],
\end{equation*}
where $\mathcal{N}$ is a normalization factor. Note that although the coefficient associated with the states with double occupancies is small, it does not vanish.}. (3) Half-filling always occurs at $\mu=U/2$ independently of $t$, so the target density remains constant during all stages of the ramp.

Specific details of the ramps used in the study are presented in Appendix~\ref{App::Ramps}.

\subsection{Exact diagonalization and density matrix renormalization group}

In order to benchmark our results we compare against numerically exact methods: ED, and the DMRG. 
We perform ED on $N=6,8,10$ chains and DMRG on $N=20$ and $100$ chains as well as the $N=4\times 4$ lattice using the iTensor package~\cite{ITensor} with an adaptable bond dimension and 30 sweeps, keeping the error for the energy below $10^{-8}$.

\section{Results}\label{sec::Results}

For the results presented in this section, unless otherwise specified, we have used $n_h =100$ hidden units, a learning rate of $\ell_r = 0.001$, and generate $N_s = 1,000$ samples for open-boundary conditions per training step. When pre-training is performed, we do it over the first $300$ training steps, using $10,000$ samples. 

For every case study, we run the training using at least 25 random initial parameters of the neural network. We call these \textit{realizations}. For all the observables $\mathcal{O}$ considered, we compute the relative error as $\mathrm{RE} = \vert 1 - \expect{\mathcal{O}}/\mathcal{O}_{GS} \vert$, where $\expect{\mathcal{O}}$ is either the value of the observable at each training step or the average of the observable over the last 100 training steps as specified, and $\mathcal{O}_{GS}$ is the exact value of the observable obtained using ED (or DMRG where explicitly indicated). We also define the \textit{best realization} as the realization with the smallest relative error in energy.

In the following, we present results for different training schemes which we refer to as \textit{methods}.
These are labeled as:
\begin{itemize}
    \itemsep0em
    \item \textbf{VMC:} Minimize the energy only. 
    \item \textbf{H VMC:} (Hybrid VMC) Perform pre-training, followed by energy minimization.
    \item \textbf{Ramp:} Perform the training while ramping $t$. The tunneling rate is initialized from a large value $t_i>t$ and decreases exponentially. The tunneling is set to its final value $t$ at 51,000 training steps.  
    \item \textbf{H Ramp:} (Hybrid Ramp) Perform pre-training, followed by training while ramping $t$, which is set to its final value at 51,000 training steps. We use the final value of $t$ also in the pre-training stage to evaluate $E$. After pre-training is finished, the tunneling rate is then set to the value obtained with the Ramp after 300 training steps (which is approximately $1.55t$). For the remainder of the realization, the tunneling rate is then updated using the Ramp.
\end{itemize}

\subsection{One-dimensional Hubbard chain}

\begin{figure*}[htbp!]
\includegraphics[width=0.9\linewidth]{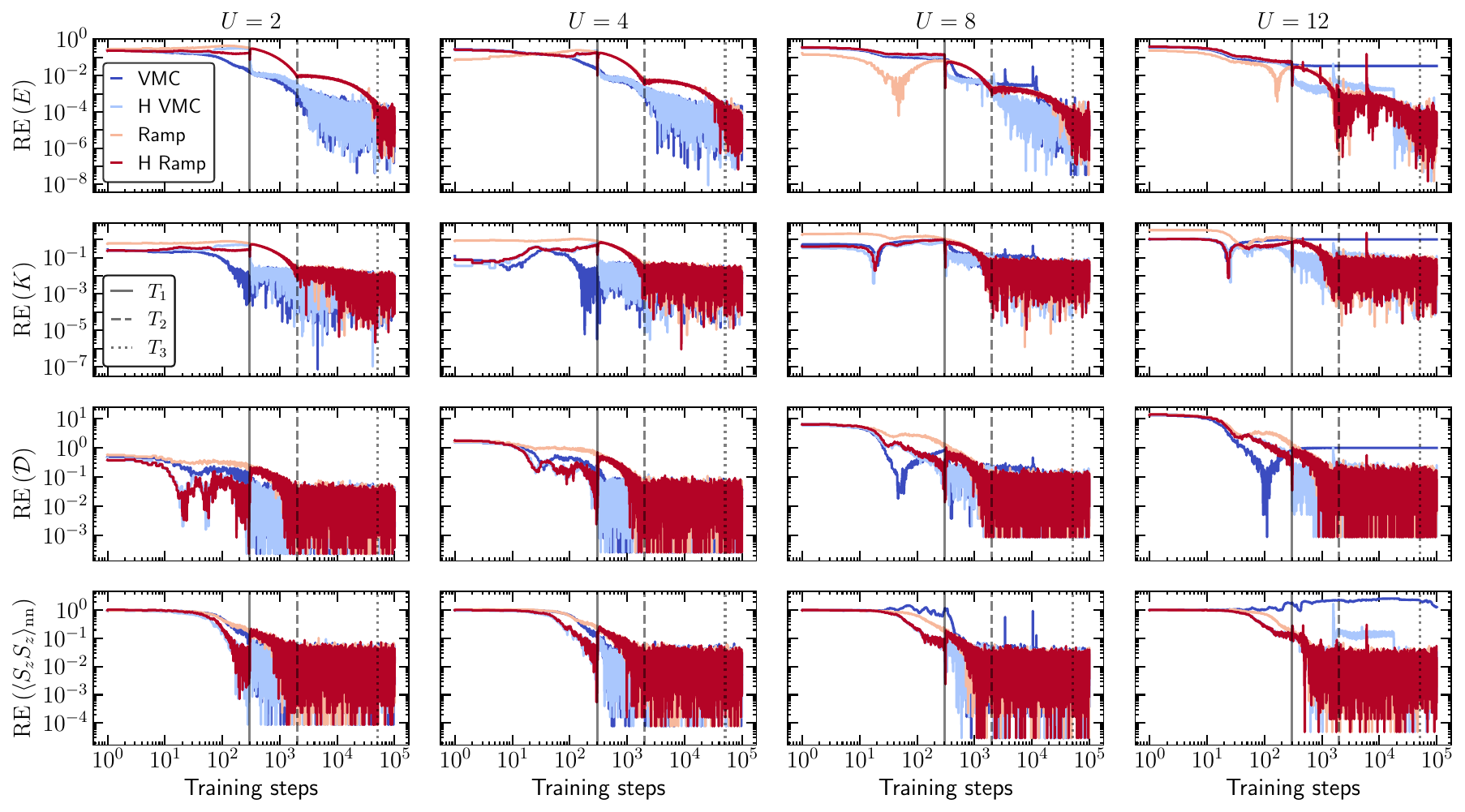}
\caption{Relative errors for the half-filled FHM as a function of training steps for different methods on a 10-site chain. Columns correspond to $U=2,4,8,12$ and rows to the relative errors in energy, kinetic energy, double occupancy, and nearest-neighbor spin-spin correlation function. Results are presented for the best random seed for all methods. Vertical lines at $T_1 = 300$ (solid), $T_2 = 2,000$ (dashed) and $T_3 = 51,000$ (dotted) training steps corresponds to pre-training and ramping stages. Sharp jumps in the H Ramp curves (red) at $T_1$ are because the tunneling rate of the Hamiltonian is changed here from $t$ to approximately $1.55t$.}
\label{fig::Rel_errors} 
\end{figure*}

In Fig.~\ref{fig::Rel_errors}, we present the relative errors of $E, K,  \mathcal{D}$, and $\expect{S_zS_z}_{\mathrm{nn}}$ as functions of training steps for $U=2,4,8,12$ on a 10-site chain at half-filling for the best realization of the RNN. In this figure, the relative errors are calculated for the expectation values of the observables at each training step and no averaging over training steps is performed.

For $U\leq 8$, all methods are able to get converged results, and their relative errors are consistent with each other. However, for $U=12$, VMC alone is unable to do so, and quickly gets stuck in the atomic limit solution ($U/t \to \infty$) where double occupancies are strongly suppressed. 

It is important to note that, for the converged results, the relative error in $E$ is approximately constant for all values of $U$ and falls within the range $(10^{-7}, 10^{-3})$. In contrast, the relative errors in $K$ and $\mathcal{D}$ worsen as the interaction strength is increased. This is evidenced by the rise in the lower and upper bounds of the ranges containing these relative errors [for example, for $K$ at $U=2$ these are $(10^{-6},10^{-2})$, but increase to $(10^{-5},10^{-1})$ at $U=12$]. This is indicative of the fact that the errors in $K$ and $\mathcal{D}$ are correlated, similarly to what is also observed in quantum Monte Carlo simulations. In contrast, the relative error of $\expect{S_zS_z}_{\mathrm{nn}}$ rapidly converges and also remains mostly constant for all $U$'s considered. It is worth noting that the lower bounds in the relative errors for $\mathcal{D}$ and $\expect{S_zS_z}_{\mathrm{nn}}$ are flat, whereas those for $K$ and $E$ continue to decrease and exhibit oscillations as a function of the training step for longer. We speculate that such difference in behavior may be traced back to the requirement of the knowledge of phases in calculating the latter quantities and the fact that $\phi_\lambda(\sigma)$ may continue to be learned after $p_\lambda(\sigma)$ are converged.

\begin{figure*}[hbtp!]
\includegraphics[width=0.9\linewidth]{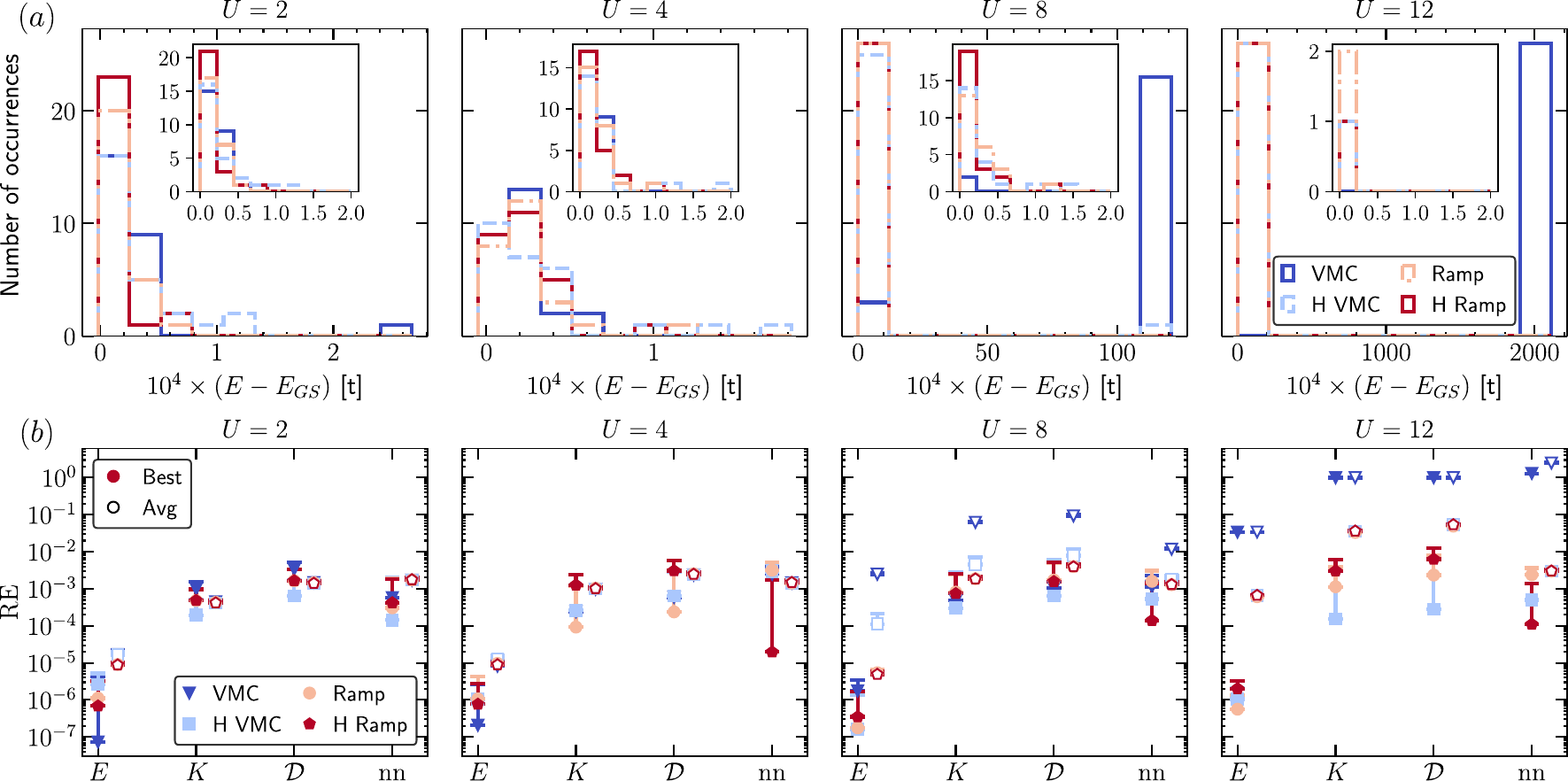}
\caption{(a) Energy histograms for the half-filled FHM for different methods on a 10-site chain. Columns correspond to $U=2,4,8,12$. Results are obtained for 26 different initial random seeds, and the energies are reported as the average of the last 100 training steps for each random initial condition. The insets replot the data in the same zoomed-in range for all $U$ to aid the comparison.
(b) Relative errors for the half-filled FHM for different methods on a 10-site chain. Columns correspond to $U=2,4,8,12$. Solid markers correspond to the best realization. Error bars are only presented for the upper bound and correspond to the mean of the relative errors obtained using $\expect{\mathcal{O}} \pm \sigma_\mathcal{O}$, where $\sigma_\mathcal{O}$ is the standard error of the mean (s.e.m.). 
Open markers correspond to the average of the relative errors obtained using 26 different random seeds and their error bars are the s.e.m. of these different realizations.}\label{fig::energy_hist} 
\end{figure*}

\begin{figure*}[htbp!]
\includegraphics[width=0.9\linewidth]{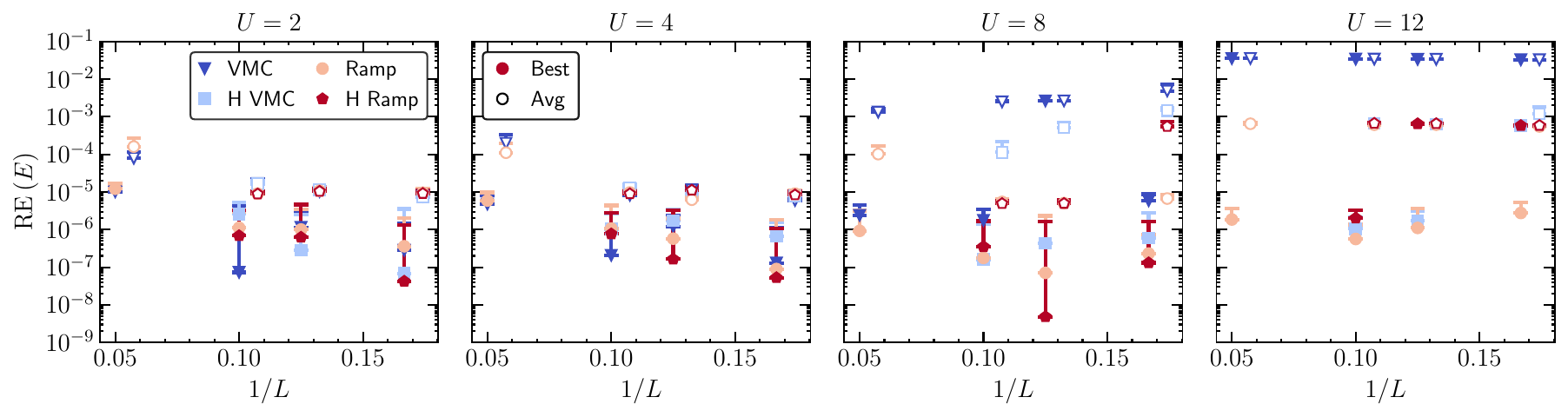}
\caption{Relative errors of the energy for the half-filled FHM for different methods as a function of system size. Columns correspond to $U=2,4,8,12$. Solid and open markers are the same as in Fig.~\ref{fig::energy_hist}(b).
}\label{fig::RE_system} 
\end{figure*}

Although almost all methods (except for VMC alone for $U=12$) are able to yield a converged RNN wavefunction with comparable relative errors ($\lesssim 10^{-6}$ in $E$ and $\lesssim 10^{-3}$ in $\expect{S_zS_z}_{\mathrm{nn}}$), an extensive search over initial random configurations may be needed to find such lowest energy configurations for some of these methods. To reveal that, in Fig.~\ref{fig::energy_hist}(a) we present histograms of the lowest energies obtained with the different methods, while in Fig.~\ref{fig::energy_hist}(b) we compare the relative errors in the observables between the best realization (solid markers) and those obtained by averaging 26 different realizations (open markers) for each method. 

At $U=2$, all methods have a large weight in states with energies very close to the ground state [see Fig.~\ref{fig::energy_hist}(a)]. However, performing VMC alone produces the least number of realizations close to the exact ground state energy. While pre-training followed by VMC (H VMC) produces better realizations, the two ramps yield a larger number of realizations closer to the exact ground state and with shorter tails. When comparing the relative errors of the different methods for $U=2$ in Fig.~\ref{fig::energy_hist}(b), we observe that except for the energy, the best realization (as judged by the relative error in energy) and the averaged data for all methods are consistent with each other within error bars. 

As the interaction strength is increased to $U=4$, the histograms exhibit a similar profile as the one observed for $U=2$, but for all methods the tails have grown, which increases the relative error of the averaged results as illustrated in Fig.~\ref{fig::energy_hist}(b). Furthermore, the results obtained using the Ramp, H VMC, and the hybrid ramp (H Ramp) yield the lowest relative errors for the kinetic energy, double occupancy and spin-spin correlation functions.

These findings are more evident when analyzing the $U=8$ results. At this interaction strength, most of the realizations using VMC alone get stuck in higher energy configurations, and only two realizations are able to get close to the ground state [see Fig.~\ref{fig::energy_hist}(a)]. On the other hand, the rest of the methods yield realizations with energies very close to the ground state, where the H Ramp method yields the largest number of realizations close to the ground state and a shorter tail for the histogram. Their relative errors in Fig.~\ref{fig::energy_hist}(b) highlight that for the best realization, hybrid VMC and the hybrid ramp yield the best results, but the ramps surpass all other methods if the number of realizations is limited.

Finally, for $U=12$, VMC alone is unable to converge to the ground state as it gets stuck in the
atomic limit solution where no doublons are present at half-filling. While the rest of the methods are capable of producing one or two converged realizations [see inset in Fig.~\ref{fig::energy_hist}(a)], the majority of the realizations exhibit an energy difference with respect to the ground state energy of $\sim 4 \times 10^{-3}$ (results lie outside the range plotted in the inset). This larger energy difference is reflected in larger relative errors in Fig.~\ref{fig::energy_hist}(b) for the open markers. Such struggle to produce converged realizations for random initial parameters is evidenced by the relative errors in energy: while for the best realization this number is $\sim 10^{-6}$, for the average over many realizations, it is roughly three orders of magnitude larger.

To summarize this section, we find that for all the interaction strengths considered, the best realizations for each method after $10^5$ training steps are consistent with each other within error bars. However, any of the ramping methods yields the largest number of runs of the lowest relative errors among the different realizations. This means that ramping the Hubbard parameters can achieve better converged results than VMC or H VMC alone.

An important question for numerical methods is how the amount of computational resources required to obtained converged results scales with the system size. In Fig.~\ref{fig::RE_system} we demonstrate that for a fixed number of hidden units, number of samples and training steps, results on $L=6,8,10,20$-site chains yield relative errors for the energy that are more or less flat within error bars (except for some slight upward trend that can be observed for $U\leq4$ when increasing the system size). This is indicative that, if the number of realizations is limited, the computational cost to keep the relative error fixed as the system size is increased beyond $L=20$ will likely scale at most linearly with system size. The observed behavior may be unique to 1D and not hold in higher dimensions, where one can in general expect a functional dependence of $n_h$ on $N$ to keep the error fixed.
In Appendix~\ref{App::Hubbard} we present results for the 1D FHM also as functions of the system size and the interaction strength, where we demonstrate they follow the expected trends as functions of these parameters. 

So far, we have benchmarked the RNNs' performance in studying the 1D FHM using various training schemes. It is worth noting that while many one-dimensional problems can be studied using efficient techniques, such as matrix product states (MPS) or the DMRG, RNNs present unique advantages over these approaches. For instance, dispersion relations are not straightfoward to calculate with MPS, but can be calculated using RNNs as demonstrated in Ref.~\cite{h_lange_23}. In order to accomplish this, the model is trained to represent the ground state initially, and then a constraint in the loss function is activated, forcing the system to reach a higher energy state with the corresponding target momentum.

\subsection{Two-dimensional square lattice}

\begin{figure}[tbp!]
\includegraphics[width=0.75\linewidth]{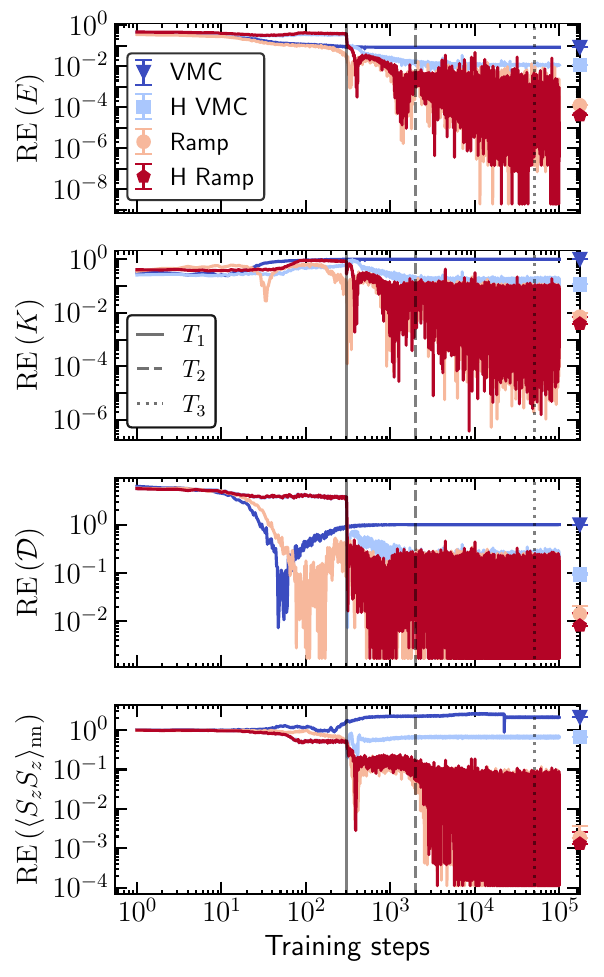}
\caption{Relative errors for the half-filled FHM as a function of training steps for different methods on a $3 \times 2$ system at $U=8$. Rows  correspond to the relative errors in energy, kinetic energy, double occupancy, and nearest-neighbor spin-spin correlation function. Results are presented for the best random seed for all methods. Light vertical lines at $T_1 = 300$ (solid), $T_2 = 2000$ (dashed) and $T_3 = 51000$ (dotted) training steps corresponds to pre-training and ramping stages.
Markers on the right axis correspond to averages of the observables for the last 100 training steps and error bars correspond to the s.e.m. }\label{fig::2x3} 
\end{figure}

Simulations of the FHM, to obtain even the basic properties, are the most challenging in dimensions higher than one and away from half-filling. For these reasons, we also explore the capabilities of RNNs with different training schemes in two dimensional systems, $3\times 2$ and $4\times4$, for $U=8$ at half-filling. In Fig.~\ref{fig::2x3} we present results for the $3 \times 2$ system using the same methods and architecture details as was done for the 1D system. In the 2D case, the benefits of using the ramps are even more evident. For the results presented in Fig.~\ref{fig::2x3}, the ramping methods achieve relative errors that are one or two orders of magnitude smaller than the hybrid VMC for all observables, thus highlighting the benefits of our physics-driven method.

\begin{figure}[tbp!]
\includegraphics[width=\linewidth]{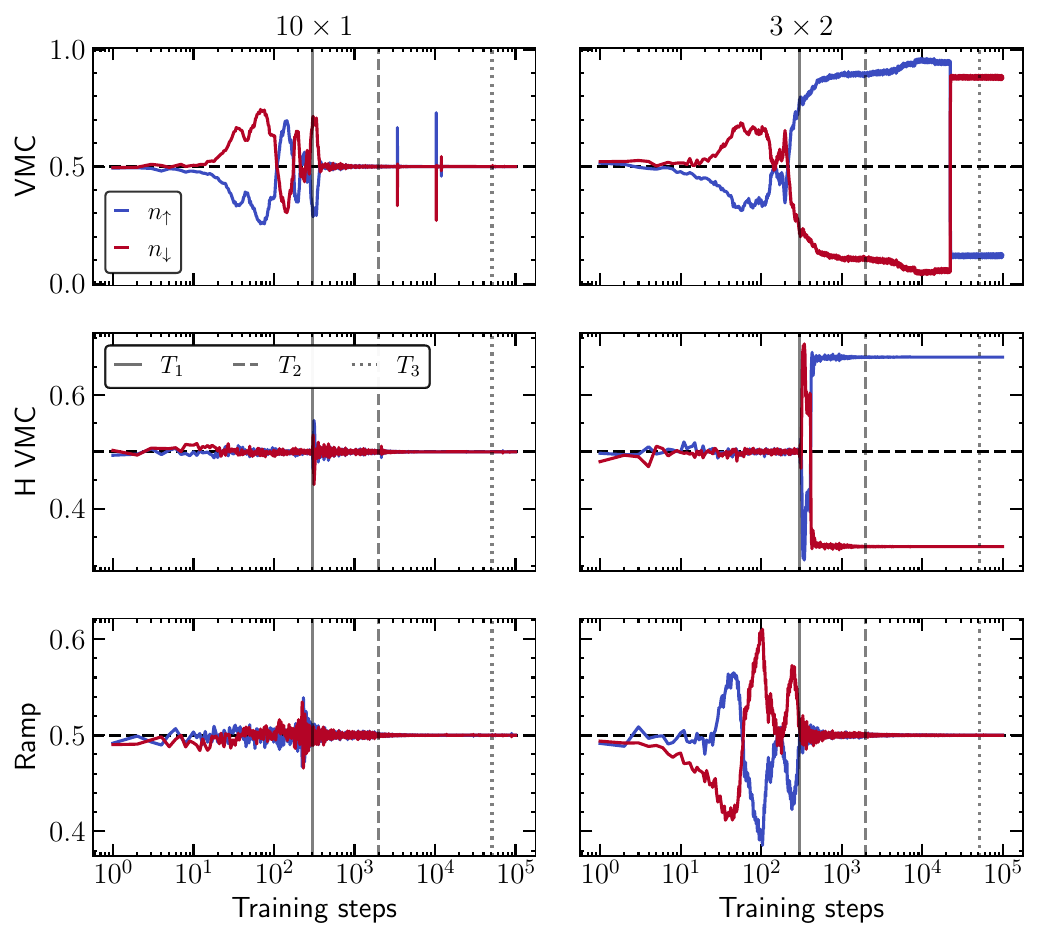}
\caption{Spin populations of the half-filled FHM as a function of training steps for different methods on a 10-site chain and a $3 \times 2$ system at $U=8$. Rows correspond to three different methods used. Results are presented for the best random seed for all methods and system sizes. Light vertical lines are the same as in Fig.~\ref{fig::2x3}.
}\label{fig::spin_populations} 
\end{figure}

Further inspection in Fig.~\ref{fig::spin_populations} shows that in 1D, the system is able to learn the balance between spin species ($\braket{\frac{1}{N}\sum_i S_z^i}=0$) for all methods considered. Note that this symmetry is not enforced during the training. In particular, as can be seen in the left panels of Fig.~\ref{fig::spin_populations} for $L=10$, we observe that the pre-training and the ramping methods rapidly help the model to learn the correct spin populations. On the other hand, the VMC oscillates and struggles to figure that out, until eventually it converges to correct values. However, in 2D, the oscillations in the VMC method are uncontrolled and lead to a polarized sample as can be seen in the right panels of Fig.~\ref{fig::spin_populations} for a $N=2\times3$ system. What is even more surprising is that although the pre-training method favors spin balanced mixtures, when VMC is turned on after the pre-training, the model immediately tends to spin polarization. Clearly the ramping method has a major advantage as it yields the best convergence towards exact results and ensures spin-balanced mixtures.

\begin{figure}[tbp!]
\includegraphics[width=\linewidth]{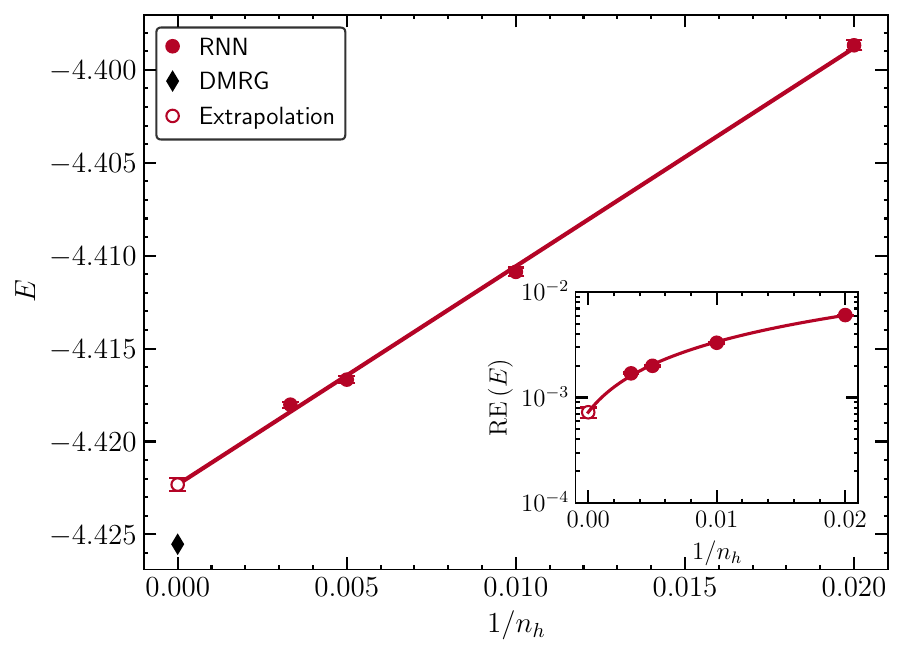}
\caption{Energy for the half-filled FHM as a function of the inverse of the number of hidden units on a $4 \times 4$ system at $U=8$. Red solid circles are RNN results, the black diamond is the DMRG result, and the solid line corresponds to a linear fit $E = E_0 +m/n_h$, where $E_0 = -4.4223 \pm 0.0004$, and $m=1.17 \pm 0.03$. The open red circle corresponds to the extrapolation to $n_h \to \infty$. Inset presents the relative error of the energy, which for the extrapolated case corresponds to $7.2\times10^{-4}$.
}\label{fig::4x4_nh} 
\end{figure}

We further evaluate the performance of the RNN using the ramping training scheme for the $4\times4$ square lattice in Fig.~\ref{fig::4x4_nh}. Specifically, we explore how the energy and the relative error in energy depend on the number of hidden units for the Ramp method. This is motivated by the fact that using $n_h=100$, the ramping method achieves a relative error in the energy of $1.2\times10^{-4}$ for the $3 \times 2$ system, but this error grows to $3.3\times10^{-3}$ for the $4 \times 4$ system. This worsening in the relative error as the system size increases emphasizes the significance of establishing the trend with $n_h$, since we know the method's computational requirements scale linearly in $n_h$. 

In Fig.~\ref{fig::4x4_nh} we observe that the energy decreases with $1/n_h$, and up to $n_h =300$, its behavior seems to be well described by a linear fit. As $n_h$ increases, the relative error in energy decreases and reaches $7.2\times10^{-4}$ for the extrapolation in the limit $n_h \to \infty$. These extrapolated results are comparable to those presented in Ref.~\cite{RobledoMoreno2022}, where the authors benchmark their relative errors against auxiliary field QMC from Ref.~\cite{Qin2016}, and illustrate the power and viability of the physics-based training scheme proposed here.

Finally, increasing the number of hidden units above 300 might change the scaling of $E$ with $n_h$ from linear to a power law, further reducing the relative error. However, these runs surpass our current computational capabilities. Lastly, it is worth mentioning that the functional dependence of $n_h$ on $N$ to keep the error fixed as a function of $N$ for the FHM in 2D remains an open question, which will be explored in future studies.

\subsection{Hatano-Nelson-Hubbard model}

In addition to the FHM in 1D and 2D, we also explore the one-dimensional Hatano-Nelson-Hubbard model (HNHM). We study this model to evaluate the applicability, power, and versatility of the method in tackling a wide range of models that do not lend themselves to traditional numerical treatments. In particular, we focus on the HNHM because (1) non-Hermitian Hamiltonians are used in the study of open quantum systems~\cite{Rotter_review}, (2) display important connections to topological materials~\cite{Nobuyuki2023,Maddi2024,Orito2023}, and (3) pose difficulties for solving with established numerical methods, despite recent efforts that have been made to explore these type of models with DMRG~\cite{ITensor,Zhong2024}.

\begin{figure}[tbp!]
\includegraphics[width=\linewidth]{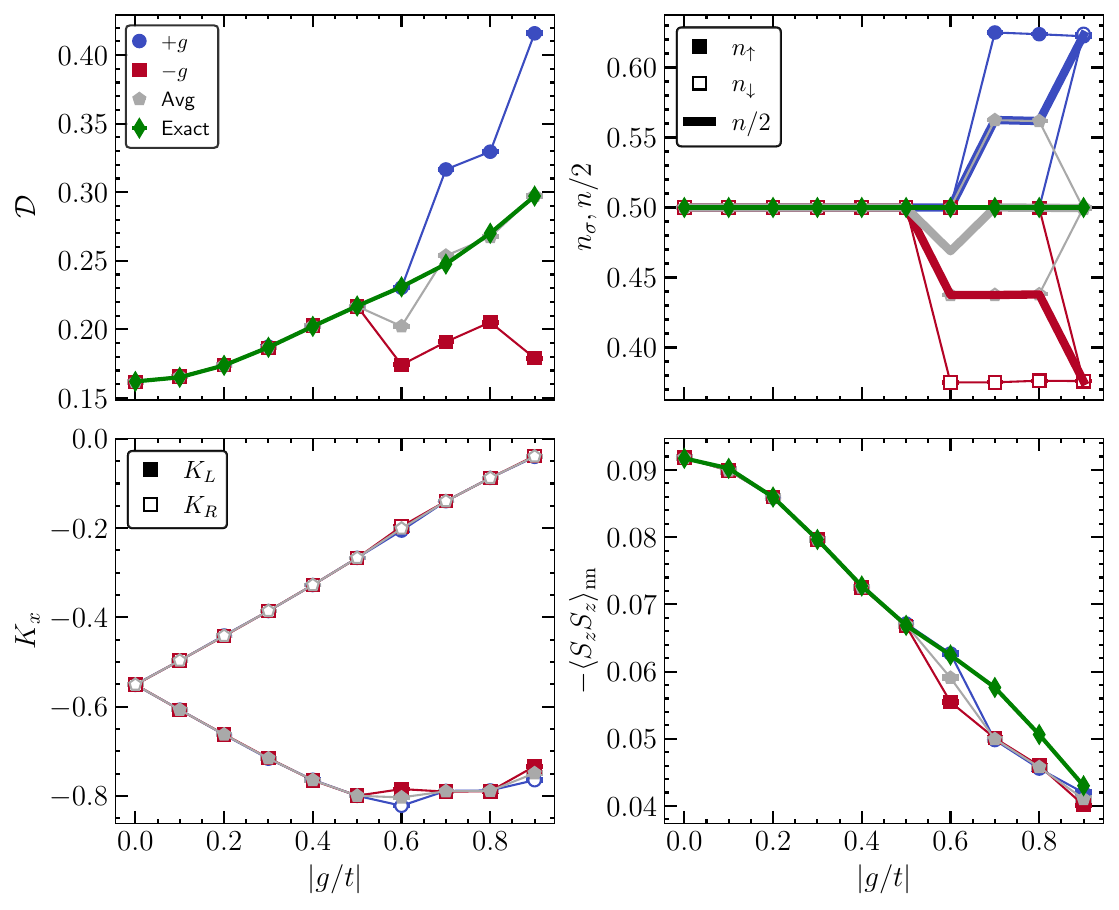}
\caption{Observables of the Hatano-Nelson-Hubbard model at $U/t=2$ and $\mu = U/2$ for the Ramp method as a function of $|g/t|$ on an 8-site chain. Results are presented for RNNs where the autoregressive sampling is performed from left to right and the favored tunneling rate is to the right (blue circles), left (red squares). Gray pentagons correspond to the average of these two independent runs, and green diamonds correspond to exact results with ED. Results are presented for the best realization, where results are obtained by averaging the observables for the last 100 training steps and error bars correspond to the s.e.m.}\label{fig::HNH_results} 
\end{figure}

Here, we focus on the HNHM with open boundary conditions. In this limit, after a gauge transformation, the model exhibits a real spectra~\cite{Nobuyuki2023}, and therefore we expect the RNN architecture to be able to accurately obtain the ground state. This is because, in current VMC implementations, the energy (or loss function) is real. Nevertheless, we find that despite the spectra being real, for sufficiently large value of $|g/t|$, the RNN fails to converge to the ground state, as shown in Fig.~\ref{fig::HNH_results} (more below).

\begin{figure*}[htbp!]
\includegraphics[width=0.9\linewidth]{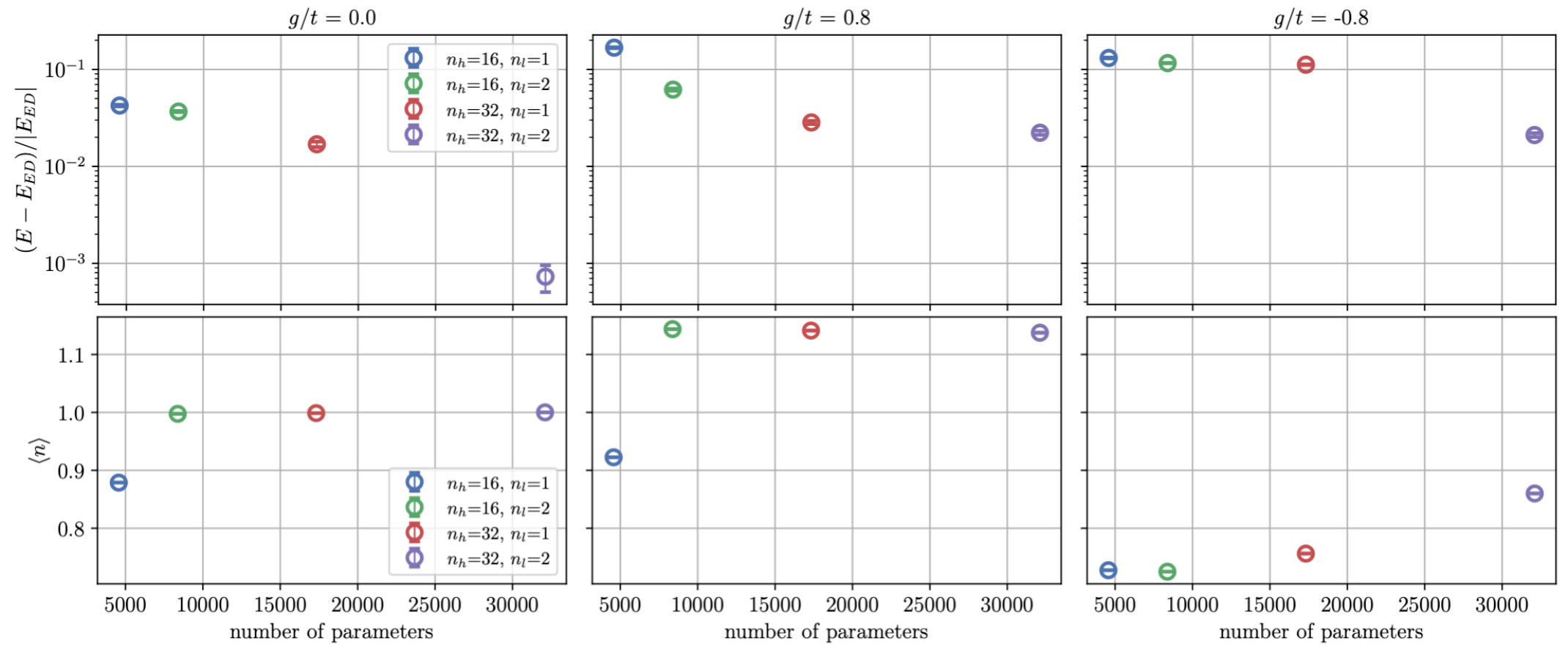}
\caption{Results obtained using a autoregressive transformer quantum state with different numbers of layers $n_l$ and embedding dimensions $n_h$. We show the relative errors (top) and the average filling $\langle n\rangle$ (bottom) at the end of the training for $g/t=0.0, \pm 0.8$ (left to right). In all calculations, $8$ attention heads are used.}\label{fig::TQSresults} 
\end{figure*}

In Fig.~\ref{fig::HNH_results}, we present results for an 8-site chain for $U/t=2$ at $\mu=U/2$ as a function of the tunneling anisotropy $|g/t|$. RNN results are obtained by performing the autoregressive sampling from left to right for two cases: (i) for $g>0$ (blue circles), which means the left-to-right tunneling is favored, and we label it as $+g$, and (ii) for $g<0$ (red squares), in which the right-to-left tunneling is now favored, and we label it as $-g$. In Fig.~\ref{fig::HNH_results} we also compare the RNN results against ED (green diamonds) for $\mathcal{D}$, $n$, and $\expect{S_zS_z}_\mathrm{nn}$.

Due to the unequal tunneling rates in the HNHM, the trends displayed by the left ($K_L$) and right ($K_R$) kinetic energies in Fig.~\ref{fig::HNH_results} as a function of $|g/t|$ are expected. At $g=0$, $K_L=K_R$. As $|g/t|$ increases towards the $g=t$ limit, the magnitude of the unfavored kinetic energy decreases, in a linear fashion, towards zero. On the other hand, as $|g/t|$ increases, the magnitude of the favored kinetic energy increases. Although the results for the favored $K_x$ suggest a linear increase in magnitude, followed by an upturn and a decrease around $|g/t|=0.6$, results for $|g/t| >0.5$ are likely unconverged as we discuss below. 

In Fig.~\ref{fig::HNH_results} comparisons against ED (green diamonds) illustrate that convergence is only achieved for $|g/t| \leq 0.5$. For larger $|g/t|$, the two realizations differ significantly: In the case of $g>0$ the system adds a particle into the chain and favors the formation of double occupancies, while in the case of $g<0$ it prefers to remove a particle from the array and disfavors the formation of double occupancies. Furthermore, the densities for the $g>0$ and $g<0$ curves behave as if they are the particle-hole transformation of each other. This behavior exposes that the relative direction of the favored tunneling rate with respect to the direction of the autoregressive sampling in the RNN plays an important role in its convergence, and that the current sampling scheme alone is not capable of recovering the PHS completely.

Motivated by these results, we then averaged the $g>0$ and $g<0$ results (gray pentagons in Fig.~\ref{fig::HNH_results}). Although these are in good agreement with the exact results for the local observables ($\mathcal{D}$ and $n$) for most values of $|g/t|$, they do not agree for the spin correlation function for $|g/t| > 0.5$, further reflecting the lack of convergence of the individual runs. These findings suggest a need for designing a different autoregressive sampling scheme. Therefore, we examined a new scheme in which we reverse the direction of the autoregressive sampling after each training step. However, we observed that the RNN's convergence did not improve using this scheme (not shown). The behavior to add or remove a particle persists. 

A similar calculation using an autoregressive version of a transformer quantum state (TQS) yields similar results. The architecture is similar to the one used in Ref. \cite{Lange2024} for spin models, but adapted to the larger local Hilbert space of the HNHM. Fig. \ref{fig::TQSresults} shows the relative errors obtained with the TQS using different numbers of layers $n_l$ and embedding dimensions $n_h$ (top) and the average filling (bottom) at the end of the training. As observed for the RNN, the accuracy decreases for $g/t\neq 0$. Furthermore, the average filling depends on the sign of $g$: It is overestimated for $g>0$ and underestimated for $g<0$, as can be seen in Fig. \ref{fig::TQSresults} (bottom) for $g/t=\pm 0.8$.

Finally, we find that the RNN's behavior of moving away from half-filling occurs for other values of $U/t$ and for different system sizes too (see Appendix~\ref{App::Hatano}). In particular we observe that the ``critical'' $g/t$ at which the system decides to add or remove a particle shifts to lower values as the system size increases and $U/t$ decreases. Such behavior is reminiscent of non-ergodicity issues in determinant QMC (DQMC), in which the method \textit{sticks} at incorrect densities at large $U$, and the incorrect density corresponds to adding or subtracting integer number of particles~\cite{Scalettar1991}. Additionally, these issues aggravate for larger systems (signatures of \textit{sticking} are present at higher temperatures for larger systems). For the FHM the relevant parameter is $U/t$. For the HNHM, we have two relevant ratios $U/(t \pm g)$. While in the isotropic limit $g=0$, $U/t=2$ corresponds to weak coupling, for $g=0.6$, $U/(t-g) = 5$, and for $g=0.9$, $U/(t-g)=20$. These findings call for further investigation into the design of RNN architectures capable of addressing non-Hermitian Hamiltonians, which will be a subject of our future studies.

\section{Conclusions}\label{sec::Conc}

In this work we utilized RNNs as a variational ansatz to access the ground state of many-body Hamiltonians. We evaluated their applicability, power, and versatility using different training schemes for the FHM in 1D and 2D, and in the 1D HNHM. 

We introduced a physically motivated method for enhancing VMC simulations that is independent of having access to experimental or numerical projective measurements. Our method is based on ramping the tunneling rate from a larger $t$ to the desired final value during the training of the RNN. 

We first benchmarked our results for the FHM against ED and DMRG on 1D chains with open boundary conditions. In this regime, (1) we observed that as system size increases, the computational cost to keep the relative error fixed will likely scale at most linearly with system size, and (2) we demonstrated that our proposed training scheme, either alone or in conjunction with pre-training, produces results that are generally better than VMC with pre-training with projective measurements.

We then applied this training scheme to the FHM in the 2D square lattice. We found that our proposed method performs significantly better than the hybrid optimization technique, achieving relative errors that are one or two orders of magnitude smaller for all observables considered in this study. Moreover, for the largest 2D system examined ($4 \times 4$) we obtained relative errors in the energy that are consistent with those obtained with NQS simulations with constrained hidden states~\cite{RobledoMoreno2022}.

Finally, our application of the method to the HNHM illustrated that further considerations need to be taken into account for the study of non-Hermitian physics with RNN. These point to interesting future studies, e.g., the use of stochastic reconfiguration for the optimization of the RNN's variational parameters~\cite{beccaQuantumMonteCarlo2017} or the use of symmetries~\cite{s_morawetz_20}. 

In addition to exploring non-Hermitian Hamiltonians, an immediate avenue for application of our method is to understand the effects of doping Mott insulators and magnetically ordered phases, which is one of the principal objectives in strongly correlated matter. In particular, the approach presented here provides a useful starting point for the exploration of the doped FHM in 1D and 2D with RNN- and TQS-based VMCs in which one may have to ramp multiple model parameters simultaneously, including $\mu$, during the training to achieve a desired filling.  

Finally, we also expect the method to perform well for the attractive FHM too since it is equivalent to the repulsive model at half filling due to the particle-hole symmetry. Studies involving the former model may provide useful insight into the convergence and scaling properties of autoregressive neural networks on large lattices in two dimensions, where QMC methods do not exhibit a sign problem away from half filling and accurate calculations of very large system sizes at low temperature are available for comparison~\cite{negativeU_scalettar1989}.

\begin{acknowledgments}
EIGP, RTS, and EK are supported by the grant DE-SC0022311, funded by the U.S. Department of Energy, Office of Science. Computing resources were supported by the Spartan high-performance computing facility at San José State University supported by the NSF under Grant No. OAC-1626645. AB and HL acknowledge the support by the Deutsche Forschungsgemeinschaft (DFG, German Research Foundation) under Germany’s Excellence Strategy—EXC-2111—390814868. HL acknowledges support by the International Max Planck Research School for Quantum Science and Technology (IMPRS-QST). RGM acknowledges support from the Natural Sciences and Engineering Research Council of Canada (NSERC) and the Perimeter Institute for Theoretical Physics. Research at Perimeter Institute is supported in part by the Government of Canada through the Department of Innovation, Science and Economic Development Canada and by the Province of Ontario through the Ministry of Economic Development, Job Creation and Trade.
\end{acknowledgments}

\appendix

\section{Details of ramps}\label{App::Ramps}

\begin{figure}[htbp!]
\includegraphics[width=0.9\linewidth]{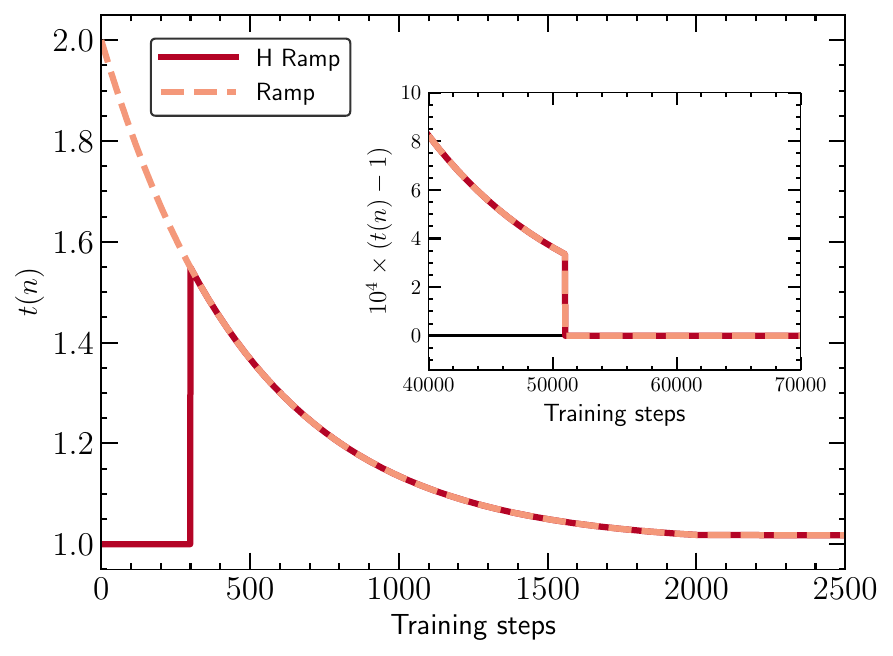}
\caption{Visualization of the tunneling ramps used in the study as a function of the number of training steps. The inset illustrates when the exponential ramp is quenched to lower the tunneling rate to the desired value of $t$.}\label{fig::ramps} 
\end{figure}

The tunneling ramps used in this study are defined as follows
\begin{align}
    t_\mathrm{Ramp}(n) &= 
    \begin{cases}
    t + (t_i-t)e^{-W n \Delta(n)}  & n < 51000 \\
    t & n \geq 51000 
    \end{cases},
    \\
    t_\mathrm{H \, Ramp}(n) &= 
    \begin{cases}
    t & n \leq 300 \\ 
    t + (t_i-t)e^{-W n \Delta(n)}  & 300 < n < 51000 \\
    t & n \geq  51000
    \end{cases},
\end{align}
where $n$ is the training step number, and 
\begin{align}
    \Delta(n) &= 
    \begin{cases}
        5 \times 10^{-4}/t & n \leq 2000 \\
        2 \times 10^{-5}/t & n > 2000
    \end{cases}.
\end{align}
We set $t_i=2t$ and $W=4t$. The ramps are illustrated in Fig.~\ref{fig::ramps} for simplicity and visualization. We also observed that different ramp parameters do not affect convergence.

\section{Further FHM details in 1D}\label{App::Hubbard}

\begin{figure*}[hbtp!]
\includegraphics[width=0.95\linewidth]{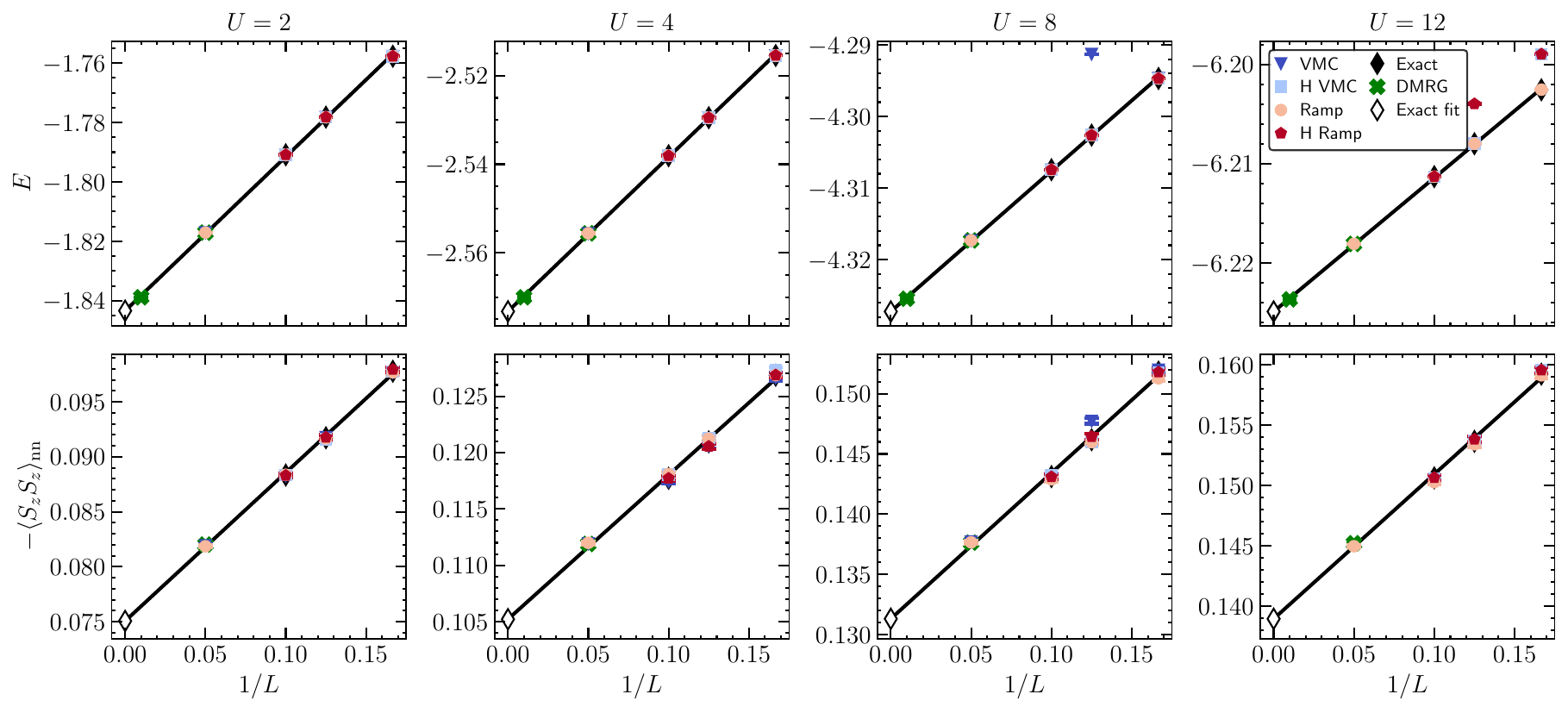}
\caption{(a) Energy (b) Nearest-neighbor spin-spin correlation for the half-filled FHM for different methods as a function of system size. Columns correspond to $U=2,4,8,12$. Results are presented for the best realization, where results are obtained by averaging the observables for the last 100 training steps and error bars correspond to the s.e.m. (except for $U=12$ for the VMC method, where results lie outside the range plotted). For $L=20$, results from DMRG are also presented for comparison (green crosses). DMRG results are also presented for $L=100$ for the ground state energy. Black solid diamonds correspond to results from ED. The black lines are linear fits to the ED and DMRG data, and the black open diamonds are the thermodynamic limit extrapolations of the linear fits.
}\label{fig::Obs_system} 
\end{figure*}

For larger system sizes where ED is not feasible we compare against DMRG and the extrapolated results from ED. These are presented in Fig.~\ref{fig::Obs_system}, where show $E$ and $\expect{S_zS_z}_\mathrm{nn}$ as a function of $1/L$. The results for the energy and the correlation function from the RNN are consistent with ED and DMRG.

\begin{figure*}[hbtp!]
\includegraphics[width=0.95\linewidth]{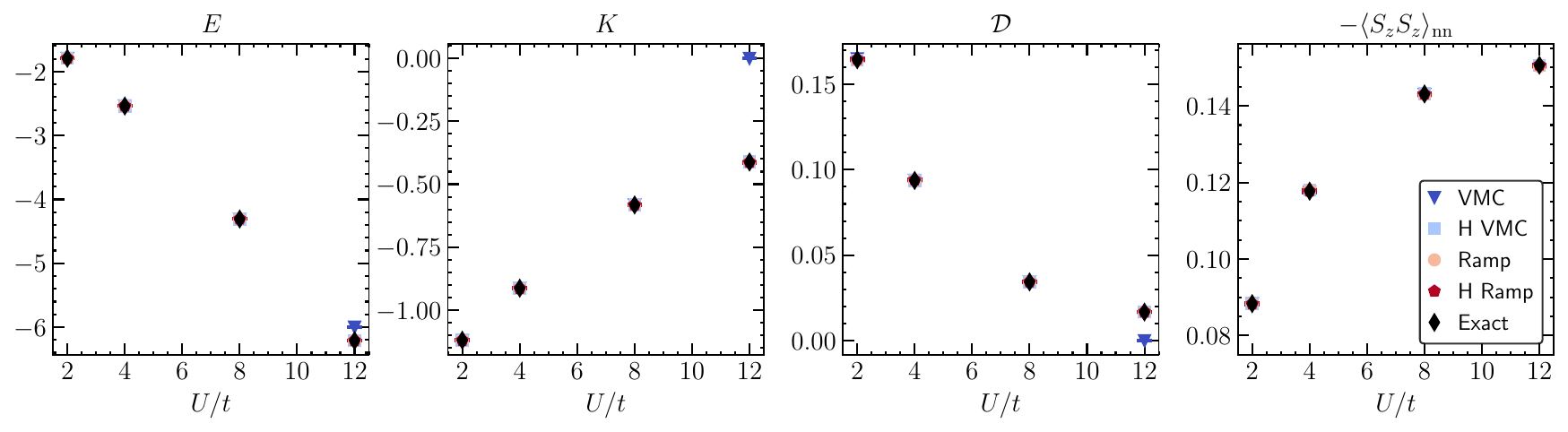}
\caption{$E$, $K$, $\mathcal{D}$, and $\expect{S_zS_z}_\mathrm{nn}$ for the half-filled FHM for different methods as a function of the interaction strength $U/t$ on a 10-site chain. Results are presented for the best realization, where results are obtained by averaging the observables for the last 100 training steps and error bars correspond to the s.e.m. For the spin-spin correlation function, the VMC method marker at $U/t=12$ is not presented since results lie outside the range plotted. Black solid diamonds correspond to results from ED.
}\label{fig::Obs_system_vsU} 
\end{figure*}

In addition, in Fig.~\ref{fig::Obs_system_vsU} we present results as a function of $U$. The behaviors are as expected, as $U$ increases, the ground state energy grows in magnitude, and both the kinetic energy and the number of double occupancies decrease in magnitude. On the contrary, the antiferromagnetic nearest-neighbor spin correlation function increases as $U$ increases before the expected decline at larger $U$. In all cases, for converged results, there is good agreement with the exact results. 

\section{More results for the HNHM}\label{App::Hatano}

\begin{figure}[hbtp!]
\includegraphics[width=\linewidth]{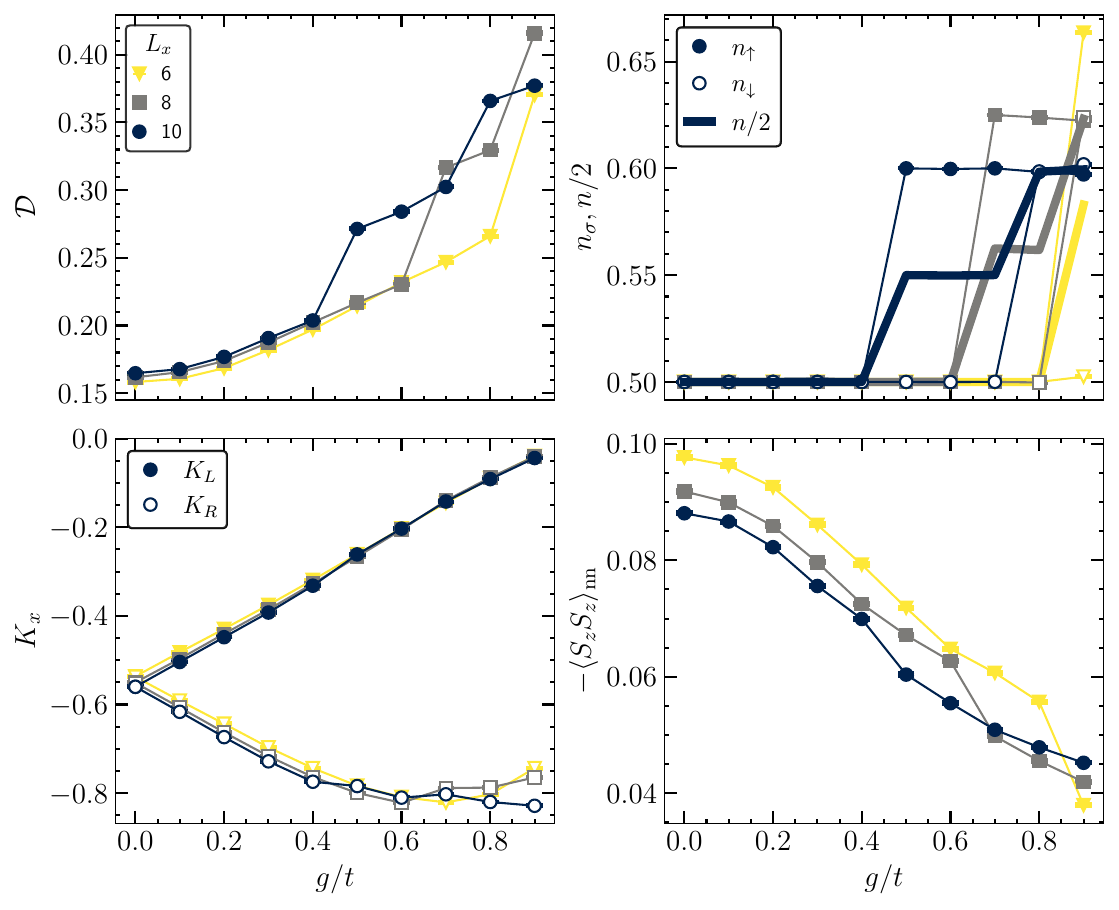}
\caption{Observables of the Hatano-Nelson-Hubbard model at $U=2t$ and $\mu = U/2$ for the Ramp method as a function of $g/t$ for different sizes $L_x$. Results are presented for the best realization, where results are obtained by averaging the observables for the last 100 training steps and error bars correspond to the s.e.m.}\label{fig::HNH_results_Lx} 
\end{figure}

\begin{figure}[hbtp!]
\includegraphics[width=\linewidth]{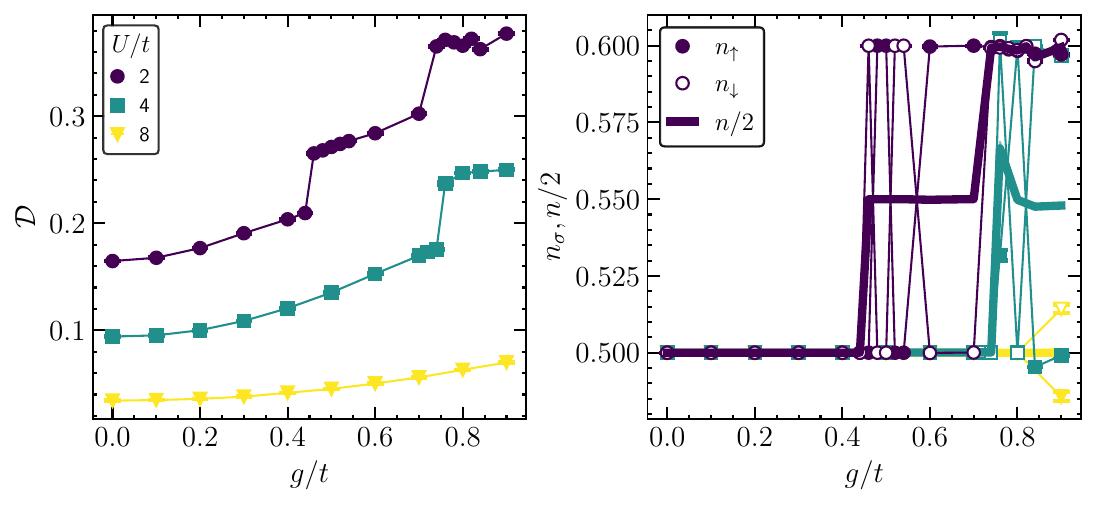}
\caption{Observables of the Hatano-Nelson-Hubbard model at $\mu = U/2$ for the Ramp method as a function of $g/t$ and different values of $U/t$ in a $L=10$ site chain. Results are presented for the best realization, where results are obtained by averaging the observables for the last 100 training steps and error bars correspond to the s.e.m.}\label{fig::HNH_results_U} 
\end{figure}

In Figs.~\ref{fig::HNH_results_Lx} and \ref{fig::HNH_results_U} we present results for the HNHM as for different system sizes and interaction strengths, respectively. Fig.~\ref{fig::HNH_results_Lx} illustrates that at fixed interaction strength, as the system size increases, the ``critical'' $g/t$ at which the system moves away from half-filling shifts to lower values. Furthermore, Fig.~\ref{fig::HNH_results_U} shows that the value of the ``critical'' $g/t$ also shifts to lower values as $U/t$ decreases.

\bibliography{NQS}
\end{document}